\documentclass[a4paper]{article}

\usepackage[margin=25mm]{geometry}
\usepackage[colorlinks, allcolors=blue]{hyperref}
\usepackage{enumitem}
\usepackage{graphicx}
\usepackage{booktabs}
\usepackage{authblk}

\usepackage[capitalize, noabbrev]{cleveref}
\usepackage[backend=biber, style=numeric]{biblatex}
\title{Cryptographically Secure Agent-to-Agent Communication in~Multi-Agent~Systems}

\author{Karol Budzyn, Adrian Brodzik, Mateusz Szczepkowski\thanks{Author emails: karol.budzyn@orange.com, adrian.brodzik@orange.com, mateusz.szczepkowski@orange.com}}
\affil{Orange Innovation Poland}

\date{}

\begin{document}

\maketitle

\begin{abstract}
Multi-agent systems (MAS) comprise AI agents that collaborate or compete to perform complex, distributed tasks. Their dynamic interactions create opportunities for threat actors to target and impersonate other agents, intercept messages, inject malicious prompts, and exfiltrate sensitive data while simultaneously complicating the detection of anomalous behavior through conventional monitoring. This paper surveys threats and cryptographic challenges in agent-to-agent communication and presents a zero-trust architecture integrating identity, transport security, message integrity, and monitoring. The architecture establishes agent identities via decentralized identifiers and task-scoped verifiable credentials backed by a permissioned ledger, enforcing least-privilege authorization. For transport security, it provides mutually authenticated, crypto-agile TLS with hybrid post-quantum key establishment, and evaluates KEMTLS with pre-distributed public keys to reduce handshake overhead. Signed CBOR envelopes preserve payload integrity across multi-hop exchanges, while zero-knowledge proofs verify declared task constraints without exposing underlying inputs. Context-aware monitoring of inter-agent messages, tool activity, and execution telemetry informs the setting of risk-based trust thresholds and supports the detection of behavioral anomalies. Together, these mechanisms provide an integrated framework for securing and auditing agent communication, with implications for other agentic protocols and prospective 6G semantic communication.
\end{abstract}

\vspace{2em}

\section{Introduction}
\label{sec:introduction}

Multi-agent systems (MAS) are defined as networks of artificial intelligence (AI) agents specialized to solve complex, distributed problems, which a single agent may not be able to solve or do efficiently~\cite{weiss1999multiagent}. Depending on the task, agents may pursue individual or shared objectives, cooperate or compete, and divide tasks according to their roles. They may also be organized through centralized, hierarchical, or decentralized architectures. A typical MAS comprises: agents, environments, interactions, and communication~\cite{maswebsite}. Accordingly, a system’s security depends not only on the properties of its individual agents but also on their coordination and information exchange, particularly where these processes cross trust boundaries~\cite{dewitt2026open}. Because MAS applications vary in scope, workflow, and architecture, their security and governance requirements are context-dependent. This variability is especially relevant as businesses, including telecommunications operators, increasingly explore MAS technologies, which are expected to play a significant role in sixth-generation networks (6G)~\cite{zhang2026semantic, pennanen20256g, ferrag2026agents}. Such networks are anticipated to support highly dynamic environments with traffic from autonomous vehicles, drone swarms, robots, and other intelligent devices, creating new challenges for security and governance.

This study reviews and merges existing approaches to securing MAS, with particular focus on agent-to-agent communication. Our work draws on state of the art literature, threat analysis reports, and empirical observations. MAS may be implemented as single-process applications or deployed over local or public networks, so the relevance of particular security measures depends on the deployment context. Some measures may therefore be redundant or unsuitable for a given system. It is assumed that agents communicate over local or public networks, where sensitive information may be sent over multiple channels and exchanged among parties with differing levels of trust. The study aims to answer the following research questions:

\begin{enumerate}[label=\textbf{RQ\arabic*:}, align=left, leftmargin=*]
    \item What cryptographic challenges arise in securing multi-agent system communication?
    \item Which cryptographic mechanisms are suitable in addressing these challenges?
    \item How can these techniques be applied to enhance the security of multi-agent system communication?
\end{enumerate}

The rest of the paper is structured as follows. \cref{sec:related} reviews the challenges faced by MAS, defines the scope of the paper, and explains why agent-to-agent communication protocols require cryptographic solutions. \cref{sec:vulnerabilities} describes common attacks on MAS, identifies the key security problems that must be addressed when standardizing agent-to-agent communication protocols, while discussing potential solutions. \cref{sec:commarchitecture} proposes a combined approach to cryptographically secure agent-to-agent communication, as well as a multi-layer approach in AI monitoring for the purpose of explainability and auditability. \cref{sec:mcp} evaluates how MCP would benefit from secure agent-to-agent communication. \cref{sec:semmcom} discusses semantic communication, its role in 6G networks, and the roadmap for its standardization. \cref{sec:discussion} revisits questions posed in the beginning and considers constraints and edge cases that may arise between the proposed architecture and its implementation. \cref{sec:conclusion} summarizes the paper and suggests future research directions.

\section{Related Work}
\label{sec:related}

Security researchers emphasize that MAS should be compliant with data privacy and security standards, at the same time they should be scalable, interoperable, decentralized and efficient~\cite{karim2025aiagents}. This challenges system engineers to strike the right balance when applying the security-by-design principle, adopting a shift-left development life-cycle, and generally implementing adequate security policies and procedures. On one hand, MAS create huge potential by means of automating monotonous or complex workloads. On the other hand, they may introduce new vulnerabilities, which combined with the power of the systems, can lead to serious business damage, such as data leaks and denial of wallet~\cite{dewitt2026open}. To effectively mitigate the risk, we should minimize the attack surface, by implementing defense-in-depth strategy. This means implementing security solutions across multiple levels, from checking LLMs to monitoring AI agents' actions.

Another challenging field is communication and explainability. Current agent-to-agent communication protocols employ different strategies to address the CIA and AAA triads~\cite{louck2026security, huang2025agentic, dewitt2026open}. What is more, considering PQC transition deadlines, it is important to build crypto-agile communication protocols that will be suitable for agent-to-agent traffic. We evaluate their strong and weak sides, to develop solutions that synthesize them. Characteristic recurrent workflow like task delegation and orchestration in MAS, without any verification lets the system gradually lose certainty that delegated AI agents are actually working on the problem the user had defined, and understood the task correctly. This can lead to burning a lot of tokens, scope creep, and waste of user's time. To answer this problem, we implement cryptographic and application layer solutions which let us verify agents' intensions and work, with integrity of the primary user context. Multi-agent systems need transparency and auditability. Aggregating rich process data, logs, and using them for monitoring and optimization of the system is a step into explainable AI, it corrects how the system works and can prevent many errors and attacks.

\section{Threat and Defense Landscape}
\label{sec:vulnerabilities}

\subsection{Common Vulnerabilities and Attacks}

This section provides a brief overview of common vulnerabilities and attack vectors in MAS. LLM-powered agents use technologies such as MCP and RAG to enhance reasoning, access additional information, manage context efficiently, and support planning and task execution. While these capabilities can help agents solve complex tasks autonomously, they inherently increase the attack surface and expose MAS to vulnerabilities in the underlying models and supporting technologies. Many attacks against agentic AI have already been documented~\cite{mitreAtlas, owaspAgentic}, but their effects can be amplified in distributed settings, where malicious inputs or outputs may be propagated between agents. We group attacks by the components they target: identity and discovery mechanisms, communication, orchestration, memory, and tools~\cite{huang2025agentic}.

Agent impersonation and naming attacks are one of the main threats targeting MAS identity and discovery mechanisms. Through typo-squatting and synonymously named agents and resources, threat actors are able to manipulate the orchestrator agent's work because their text processing algorithms can classify synonymously named resources as equivalent, or better than the original ones (shadowing attacks), and thus engage with inappropriate, malicious agents or tools, creating dangerous and faulty workflows. This can as an initial vector, letting more sophisticated threats into the system.

Direct and indirect prompt injections represent another serious threats targeting MAS, leading to RAG and memory poisoning. They compromise an agent's reasoning and context through, oftentimes, indistinguishable changes or steganography applied in documents or prompts which the agent parses and evaluates. This can lead to data leaks, flooding attacks, and waste of resources, making MAS inefficient. This shows, how important it is to define appropriate input and output guardrails, authenticate and compare agents' work to the original user's intent, and to be able to employ task rollbacks.

The rug pull attack is also noteworthy in the context of MAS. An attacker initially provides a benign, useful agent which has been approved and is trusted. Later, as it operates and updates, it may begin to exhibit anomalous behavior, abuse the acquired trust and permission levels to try to perform malicious actions; the act is analogous to software CI/CD and dependency rug pulls, as well as Web3 and DeFi liquidity rug pulls. A zero-trust approach is highly recommended. One counteraction is to pin the dependency to a specific version or cryptographic hash, and only approve newer versions after thorough verification. However, static analysis is not enough to prove safety, hence it is crucial to constantly monitor and detect any suspicious actions in the systems at runtime. There are many other types of attacks targeting MAS, for a deep analysis we refer the reader to the agent-to-agent security benchmark~\cite{li2026a2asecbench}.

\subsection{Mitigation Techniques}

A range of security solutions have been proposed to mitigate the aforementioned attacks. We divide them into three key phases of threat mitigation: prevention, detection, resilience, and reaction. The first phase is very critical, as it aims to minimize the attack surface and stop malicious actions before they can be executed. This phase includes solutions like static analysis of LLMs, agent definitions, dynamic sandboxing of untrusted code, external files, and entire agent environments. Unique identities are assigned to agents with limited scope following the principle of least privilege, by limiting which resources, tools, memory, and APIs an agent can access. Apart from that, cryptographic solutions ensure data integrity and confidentiality through encryption of data at rest and in transit, including secure agent-to-agent communication channels which are further discussed in~\cref{sec:commarchitecture}.

Phase 2 can be implemented through continuous, context-aware monitoring combined with trust-threshold models that assess agent behavior and interactions. MAS are complex and nondeterministic, making existing monitoring stacks designed for software inadequate. An agent can be fully compromised while conventional system-health metrics remain normal. The critical risk is not only the compromise of an individual agent, but also the propagation and amplification of its actions or errors across the system, leading to cascading failures. For example, a user-facing agent may misinterpret a benign request and delegate an unsafe action to a more privileged agent (privilege escalation). Monitoring tools generally fall into two complementary categories: application-level and system-level. The former tracks semantic context, including prompts, inter-agent messages, tool calls, outputs, configuration, and decision traces. The latter records low-level execution events, such as process creation, file reads and writes, and inbound and outbound network activity. Without end-to-end correlation, a system-level alert could identify suspicious execution but would not be able to establish its causal provenance, i.e., determine whether the action was triggered by a legitimate user request, an adversarial prompt, or a model hallucination.

To bridge this divide there is a shift toward boundary tracing. A good example of this approach is AgentSight~\cite{zheng2025agentsight}, an observability framework that leverages the extended Berkeley Packet Filter (eBPF) to bridge the semantic gap between the two monitoring paths. These programs are verified and run securely within the Linux kernel, capturing real-time data without altering or slowing down the monitored processes. AgentSight intercepts relevant system calls and encrypted traffic at the kernel-level and extracts the actual semantic intent sent to and from external LLMs. By applying correlation algorithms, the framework can link the high-level LLM responses with the behavior they trigger. As MAS continues to mature, this kind of solution will have to scale across those distributed, highly complex systems. Concepts like automatic fault injection, real-time cost monitoring will also need to be addressed in order to ensure proper auditability and traceability. This will be an important source of operational data for trust threshold models.

The last phase focuses on a system’s ability to withstand active compromise and contain threats to minimize resulting harm, an essential capability in MAS. One safeguard is human-in-the-loop oversight, i.e., high-stakes actions require human authorization, giving operators an opportunity to stop potentially malicious or compromised behavior before it causes harm. Attacks on orchestration or reasoning can also trigger denial-of-wallet attacks by driving excessive resource use. Per-workflow budget limits can mitigate this risk by limiting runaway reasoning loops and scope creep. However, budgets that are too restrictive may disrupt legitimate workflows. Finally, regular red teaming and adversarial testing can uncover hidden vulnerabilities while exercising incident-response capabilities and helping teams refine their containment procedures.

\subsection{Telecom Perspective}

Building secure MAS is challenging, but doing so is essential to realizing the potential of distributed systems, particularly in telecommunications. MAS offer a promising approach to networks that can be continuously monitored and adapt to changing conditions. By automating management processes and accelerating fault response, they can improve operational efficiency, reduce energy consumption, and support network orchestration based on real-time demand~\cite{pennanen20256g}. For telecommunications providers, however, MAS are more than tools for incident response and infrastructure management. 6G standards emphasize close integration between network infrastructure and MAS, envisioning infrastructure as more than a data conduit, i.e., it becomes an active platform for intelligent, distributed services~\cite{ferrag2026agents}. This vision makes secure MAS deployment and agent-to-agent communication essential.

\section{Secure Communication Architecture}
\label{sec:commarchitecture}

Agent-to-agent communication is a foundational component of MAS. Unlike many conventional microservice and API interactions, which are designed around predefined interfaces and workflows, agent-to-agent communication is dynamic, nondeterministic, goal-oriented, and often transient~\cite{google2025a2a}. Securing these interactions requires more than traditional encryption, authentication, and authorization. It also calls for mechanisms to verify intent, account for workflow context, and detect suspicious or anomalous inter-agent commands~\cite{louck2026security}. Identity models must likewise accommodate non-human agents whose identities may be short-lived or change dynamically~\cite{huang2026zerotrust}. Based on these requirements, we argue that a hybrid blockchain may be suitable for supporting secure agent-to-agent communication. This section examines technologies that can contribute to the development of such protocols.

\subsection{Blockchain-Based Identity Management}

MAS shift the paradigm from static, deterministic systems towards dynamic, ephemeral, learning systems whose behavior evolves throughout their life-cycle. This poses new challenges for identity management, as traditional identity access management (IAM) was not designed for such environments. We must therefore accurately limit the scope of what an AI agent's identity is. In general, it is understood as a composition of different attributes, describing its capabilities, behavior, relationships, attestations, origin, and purpose~\cite{huang2026zerotrust}. This set depends on implementation. In zero-trust architectures, one approach to representing them relies on decentralized identifiers (DIDs) and verifiable credentials (VCs)~\cite{zou2025blocka2a}.

\begin{figure}[p]
    \centering
    \includegraphics[width=\textwidth, trim=1cm 5cm 1cm 4cm, clip]{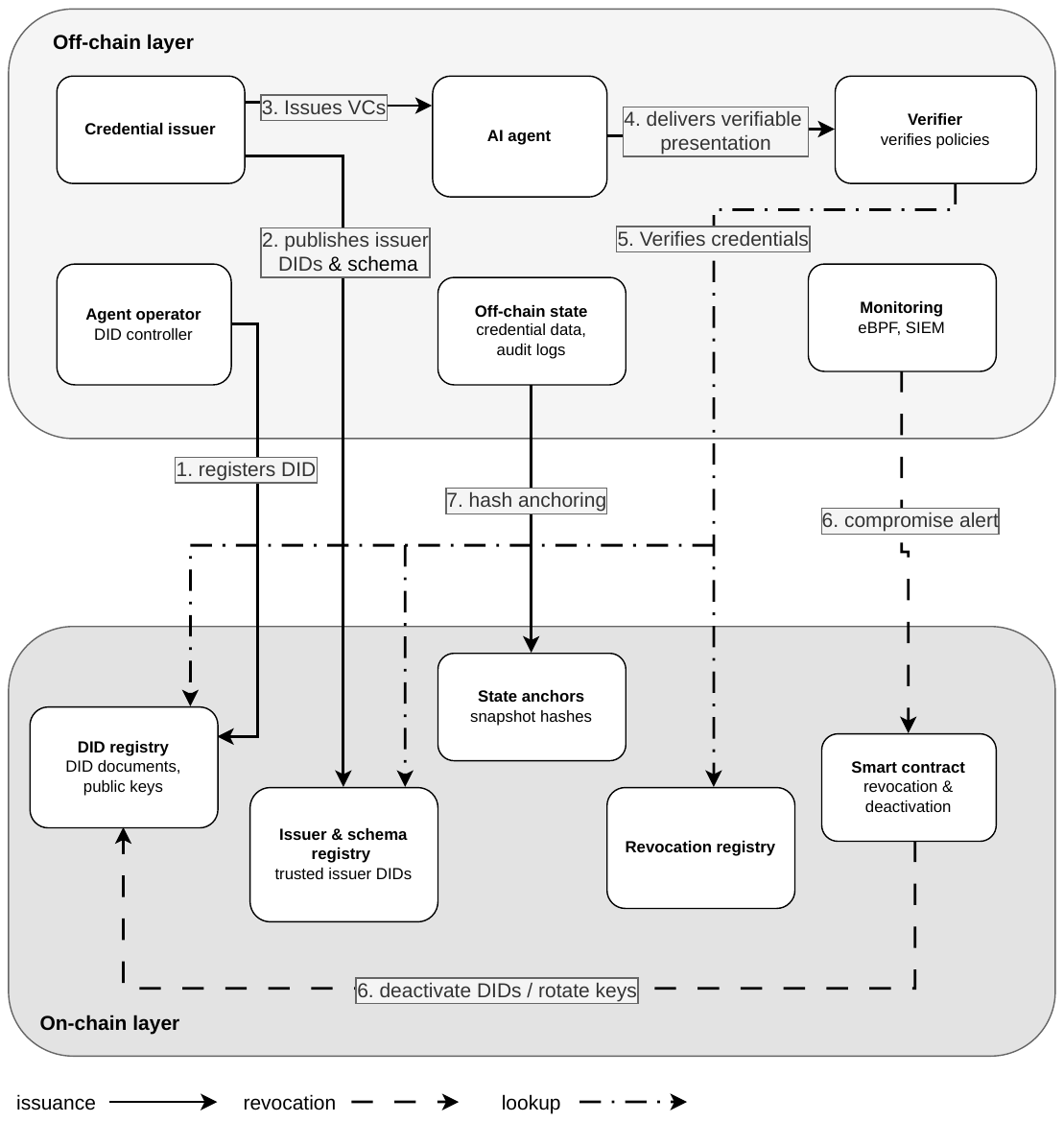}
    \caption{Identity management architecture based on a permissioned ledger.}
    \label{fig:idflowchart}
\end{figure}

DIDs provide cryptographically verifiable agent identifiers, that are globally unique and managed by the agent's owner, typically the agent's operator. DID documents contain the agent's public keys and verification methods allowing the agent to prove control over its identity. The history of DIDs should be immutable, accessible, and logged, to enable proper monitoring and auditing~\cite{neulinger2025fostering}. VCs, in turn, carry the agent's attributes and capabilities. A VC is a signed attestation issued by a trusted issuer, and held by the agent to present it to a verifier when needed~\cite{huang2026zerotrust}. The verifier check's the issuer's signature and the credential's status whether it is active, expired or revoked. VCs themselves are immutable. Dynamically changing capabilities are realized by revoking the outdated credentials and issuing the new ones. Following the just-in-time (JIT) principle, an agent gains only short-lived permissions specific to the orchestrated task that are compliant with its scope and assigned just for the runtime of it, which prevents standing permissions that can be potential attack space.

In the hybrid approach DID documents, issuer identifiers, credential schemas and revocation registers are stored and anchored on the blockchain ledger, while VCs remain held off-chain with their holders; as shown in~\cref{fig:idflowchart}. This separation keeps personal and sensitive data off the immutable ledger, while still allowing any verifier to check credential status. The off-chain state is periodically anchored on-chain through snapshots to update and archive the system's state~\cite{neulinger2025fostering}. All DIDs and VCs must be kept up-to-date, monitored, audited~\cite{karim2025aiagents}, in accordance to the least privilege principle. When a threat actor manages to compromise an agent, a monitoring system triggers a smart contract that rotates the agent's keys or immediately deactivates its DIDs and globally revokes all of its credentials through the on-chain registry~\cite{neulinger2025fostering}. A permissioned blockchain distributes keys, identities, but is not the root of trust.

\subsection{Quantum-Proof Transport Security}

Authenticated encryption in agent-to-agent communication protocols is mandatory, as even simple messages exchanged between agents can carry some valuable business information. Authenticated encryption guarantees the confidentiality and integrity of the exchanged messages. Protection against man-in-the-middle (MiTM) attacks requires both parties to be authenticated during key establishment, relying on the defined scheme. To establish a secure session, two sides need to perform mutual authentication. Typical TLS deployments authenticate the server, while the other side remains anonymous at the transport level and relying on application-level credentials such as bearer tokens, which can be stolen and replayed, while mTLS additionally authenticates the client. In agent-to-agent communication both parties are autonomous peers,
that is why mTLS should be implemented instead. This way the conversation initiating agent verifies that it speaks to the intended, authenticated interlocutor, and the receiving agent that gets the conversation request verifies the identity of the sender before any message content is processed. If the mTLS handshake goes through, a mTLS tunnel is established for quick message transfers, enabling efficient communication.

Identity frameworks like SPIFFE (secure production identity framework for everyone) constitute the root of trust, provision agents with short-lived, automatically rotated identity documents called SPIFFE Verifiable Identity Documents (SVIDs) issued by implementations such as SPIRE. The SVIDs certificates in X.509 format bind mathematically the agent's key pair to its SPIFFE ID, a URI identifying the workload. In the architecture proposed in~\cite{nirban2026zerotrust}, DIDs and VCs are vouched by SVID X.509 certificate and they are managed by smart contracts on the ledger. SVID certificates are ephemeral, this limits the exposure window of a compromised identity and once the agent's identity is revoked via blockchain registry, its SVID is no longer renewed and subsequent mTLS handshakes will fail. If we want to get an immediate effect, we need a verifying peer to check the identity status in the registry during certificate validation and implement dynamically reacting prevention solutions.

mTLS establishes a tunnel protected by authenticated encryption with associated data (AEAD), e.g., AES-256-GCM, ensuring that the exchanged payloads, such as messages encoded in the concise binary object representation (CBOR) and zero-knowledge proofs (ZKPs) can neither be read nor modified by an eavesdropper. While symmetric AEAD ciphers with 256-bit keys remain secure against harvest now decrypt later attacks (HNDL), asymmetric cryptography does not. This  motivates the adoption of post-quantum and hybrid cryptographic schemes discussed in the following section.

Both phases of the classical mTLS, i.e., authentication and key exchange, rely on asymmetric cryptography. Authentication is handled naturally with digital signatures using typically RSASSA-PSS or ECDSA, while session keys are established via ephemeral elliptic curve Diffie-Hellman (ECDHE) key exchange. The integer factorization problem used in RSA and discrete logarithm problem used in ECC will both be solved in polynomial time by a cryptographically relevant quantum computer (CRQC)~\cite{neulinger2025fostering}. This is why it is obligatory to implement and evaluate post-quantum cryptography algorithms in agent-to-agent communication by means of post-quantum mTLS, and to build crypto-agile protocols and infrastructure as post-quantum cryptography standards may dynamically change and are still evolving. However, we should not just naively replace classical algorithms with their post-quantum counterparts, for example ECDSA with ML-DSA. Comparing ML-DSA signature, which is 2420 to 4627 bytes long depending on the security level, to 64 bytes long signatures for ECDSA, we can observe a huge difference in weight. This may lead to bottlenecks in the systems, since a certificate chain carries several signatures and public keys, while the handshake may exceed the initial TCP congestion window. Additional round trips and fragmentation may occur, which is particularly costly on constrained links such as those of drones and IoT devices.

To avoid this from happening, different cryptographic approaches need to be evaluated, from hybrid to fully post quantum cryptography algorithms. The bottleneck lies in the transmission overhead rather than in computation, as ML-DSA verification is in fact faster than that of ECDSA~\cite{schemitt2025pqc}. This motivates hybrid key exchange schemes (e.g., X25519MLKEM768) with more compact signatures such as FN-DSA, certificate compression, and KEM-based authentication as in KEMTLS. KEMTLS redesigns the TLS handshake by replacing the online handshake using heavy PQC signatures~\cite{schwabe2020pqtls} with implicit authentication based on  key encapsulation mechanisms (KEMs) especially ML-KEM, for key exchange and authentication. In this flow, the responding agent's certificate would contain KEM public key rather than signature verification key, only the holder of the corresponding private key can decapsulate the shared secret and derive the session key. However, the certificate itself is still signed by a CA, so post-quantum signatures remain still in the certificate chain. This overhead is removed by KEMTLS wit pre-distributed keys (KEMTLS-PDK)~\cite{wiggers2024pqtls}, in which the initiator already knows the respondent's long-term KEM public key. A secret to this key can be encapsulated in its first sent message, which authenticates the responder within the first round trip and eliminates the transmission of the certificate. Client authentication, required for mTLS, can be performed proactively in the same flight. At the time of writing, KEM-based authentication in TLS is being standardized by the IETF as AuthKEM.

\begin{figure}[h]
    \centering
    \includegraphics[width=\textwidth, trim=1cm 5cm 1cm 4cm, clip]{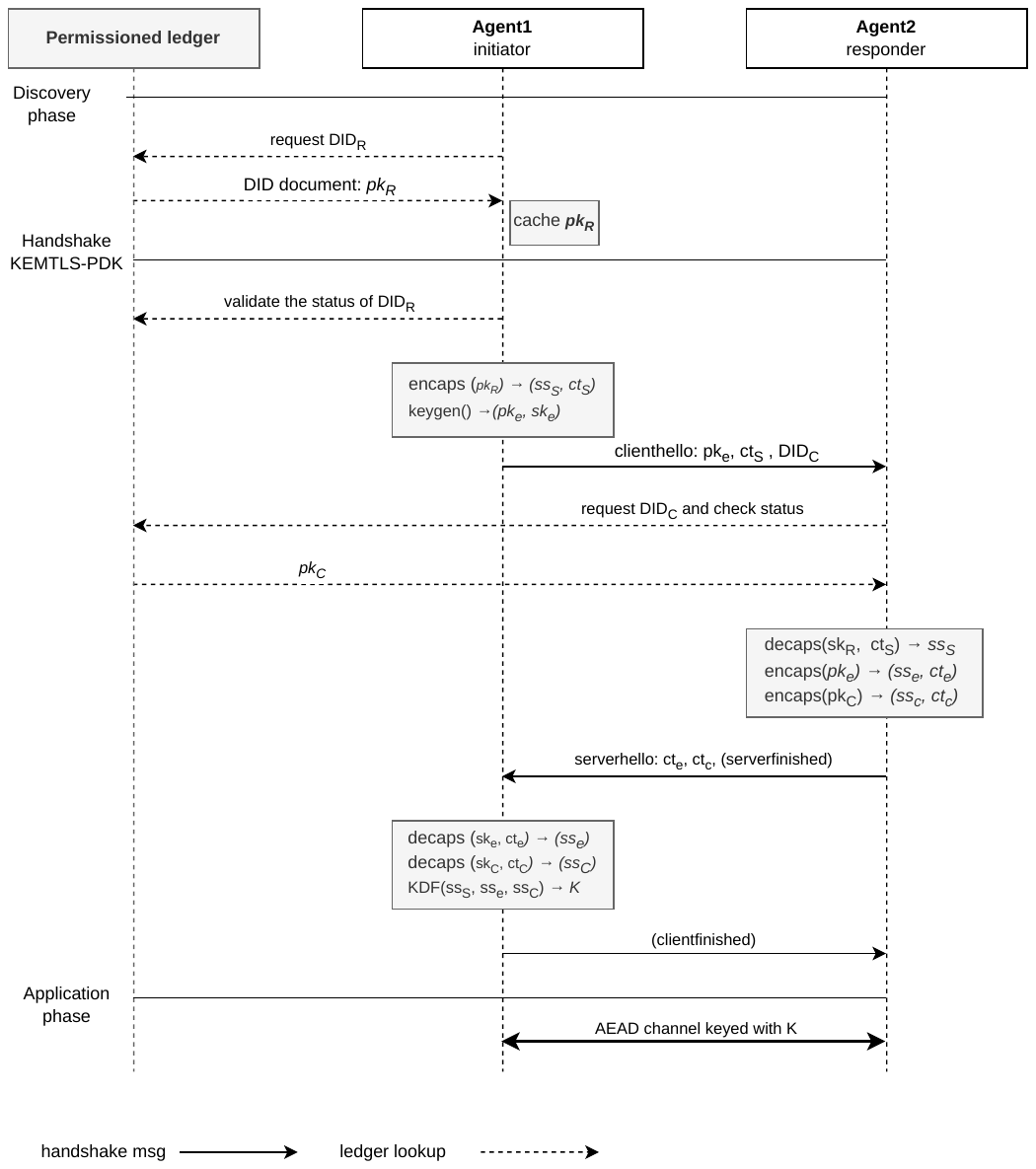}
    \caption{KEMTLS-PDK handshake with DID-based key distribution.}
    \label{fig:kemtlspdk}
\end{figure}

We propose that MAS leverage KEMTLS-PDK, as shown in~\cref{fig:kemtlspdk}, as agents in a zero-trust MAS discover each other's DIDs and permissions through a ledger-based decentralized blockchain registry, we can pre-distribute long-term ML-KEM public keys into the agents' DID documents during the discovery phase. When the orchestrator agent initiates a connection it encapsulates the shared secret to the cached public key of the responder. This lets agents to exchange authenticated data with low latency overhead, offsetting cost of post-quantum algorithms. This approach makes the ledger become the root of trust for the KEM keys, so the transactions registering and updating DID documents must themselves be secured by PQC signature schemes, otherwise a quantum adversary could substitute an agent's KEM key. What is more the initiator must verify the status of the responder's DID before using a cached key, as the key may have been rotated or the DID deactivated in the meantime.

Post-quantum cryptography algorithms are  still a relatively new, especially their implementations. Unlike well-established cryptosystems like RSA or ECC, PQC have undergone far less cryptanalytic scrutiny, as illustrated by the breaks of the SIKE and Rainbow candidates late in the NIST standardization process, and by side-channel vulnerabilities found in early ML-KEM implementations. For now, they are too immature to just start exclusively relying on them. It is preferable to implement the hybrid schemes first, before fully transitioning the system to PQC, and to follow coordinated migration road-maps. For example, we could be using post-quantum and traditional cryptography in the hybrid key exchange, which is called the PQ/T, enabled in TLS 1.3. During the TLS handshake agents perform a classical key exchange (with X25519) and the post-quantum key encapsulation (ML-KEM) simultaneously, both key shares transmitted in the same messages. Then, both resulting secrets are fed into the standard TLS 1.3 key schedule, in place of the (EC)DHE secret, from which the session keys are derived~\cite{rfc9954}.

What is more, MAS need crypto-agility, which we gain by software inventories, especially through creating cryptographic bills of materials (CBOMs). This ensures the communication stream remains secure even if one of the underlying algorithms is compromised. This way we are able to achieve a resilient, quantum-proof zero-trust perimeter, ready to safely transport the ZKP payloads and operate securely.

\subsection{Zero-Knowledge Proof Intention Verification}

In zero-trust MAS, authenticated channels alone are insufficient to ensure the integrity of the delegation process. While a mTLS tunnel between agents provides confidentiality and integrity of the exchanged messages, it offers no guarantee that the content delegated by an agent remains consistent with the task it was authorized to perform. It is mandatory, to create a system which allows the receiving agent to prove this consistency, without recalculating the task or inspecting the inner state of the sender agent.

Zero-knowledge proofs (ZKPs) are promising solution for this purpose~\cite{neulinger2025fostering}, they let a prover convince a verifier that a set of arithmetic constraints holds for some secret witness, without revealing the witness itself. The task intent is anchored in the JIT VCs issued to the agent by the task issuer. The task issuer embeds a commitment to the task policy, such as the permitted payload schema or semantic scope, into the signed credential, while the verification keys of the corresponding circuits are held in the ledger-based schema registry. When the first agent $A$ delegates a task to a second agent $B$, it attaches a succint proof, e.g, zk-SNARK, that the delegated payload satisfies the committed policy without disclosing the payload. This lets the system achieve what is called spatial-semantic zero-knowledge routing (SS-ZKR)~\cite{touheed2026sszkr}, which binds differentially private semantic intent vectors to zero-knowledge proofs of payload-schema consistency enabling routing across trust boundaries without payload decryption. The receiving agent verifies the proof against the commitment contained in the credential; for succinct proof systems, verification takes milliseconds regardless of the size of the underlying computation, whereas proof generation remains computationally expensive. If it satisfies the polynomial constraint system associated with the user's intent, the receiving agent gets a proof that the payload is compliant with the constraints encoded in the circuit, but it does not attest to the semantic correctness of the model's output.

Hallucinated or malicious results caused by threat actors will still yield a valid proof as long as it conforms to the policy. ZKPs simply limit the impact of these events rather than prevent them. However, many zk-SNARKs constructions rely on elliptic-curve pairings, which are not quantum-resistant, to answer this we can go on with hash-based systems such as zk-STARKs. This part is optional in securing MAS. ZKPs deliver proofs of correct execution for settings in which agents process confidential data that must not be disclosed, yet other parties still need assurance that the task has been performed in accordance with the agreed policy. This allows verifiers to confirm, that a specified workload was executed and its output satisfies the policy standards encoded in the registered circuit without accessing the underlying confidential data.

\subsection{Message Interfaces and Envelopes}

Agent-to-agent communication protocols often use interfaces based on client-server architecture. They implement JSON-RPC, gRPC or REST API as their standard. While these are conventional solutions, in zero-trust MAS, where each node generates and verifies ZKPs, the interface must guarantee strict determinism.

Firstly, JSON-RPC needs canonicalization before its used in ZKP schemes. When an intermediary agent parses a JSON payload it then re-serializes it to send the message further, which can mutate the original message, leading to the different hash value breaking the chain of trust~\cite{rfc8785}. REST can introduce limitations in agent-to-agent communication when interactions are long-running or inherently asynchronous. Its conventional request–response model may be less suitable for asynchronous task execution, intermediate result exchange, and event-driven agent interactions. Mixing it with the asynchronous model, in which MAS operates, can lead to huge timeout cascades and high latency of ZKP generation. Lastly, gRPC is a commonly implemented solution for these protocols, as it handles complex and autonomous interactions better than JSON or REST. gRPC is quick and lightweight, which makes it fit in environments with fluctuating network stability. Additionally, it allows for heavy orchestration logic to live in the cloud, while edge agents could seamlessly connect, report state, and fetch tasks without overwhelming local resources.

While gRPC utilizes Protobuf and secures tunnel via mTLS, it may not secure the payload across the agents. An intermediary agent could technically tamper with the Protobuf payload before sending it to the next agent, necessitating a second layer of cryptographic validation~\cite{saltzer1984endtoend}. Hence, an immutable cryptographic envelope is recommended to handle payload security, including digital signing. To achieve this, instead of relying on the transport layer, messages could utilize a binary, deterministic format such as Concise Binary Object Representation (CBOR) paired with COSE envelopes or a strictly nested Protobuf structure in gRPC~\cite{rfc9052}.

\subsection{Trust Threshold Model}

The aforementioned cryptographic mechanisms provide essential security guarantees, however, cryptographic authorization alone does not establish whether an agent operates effectively, reliably, and cost-efficiently. The proposed operational trust layer complements these mechanisms by evaluating agents' observed behavior and informing agent discovery and task routing. The model follows a zero-trust principle, i.e., an agent's suitability is assessed in context rather than inferred solely from its prior authorization. A dedicated trust-evaluation service or agent computes and updates each participant's score using indicators, such as ZKP verification success rate, latency, token consumption, economic efficiency, and telemetry from the monitoring solution described in~\cref{sec:vulnerabilities}. These indicators are assessed against relevant KPIs and other criteria, and the resulting scores are recorded in the blockchain registry.

When a task is submitted, the orchestrator determines its risk profile and selects a corresponding trust threshold. An agent may be assigned the task only if it meets the applicable threshold and satisfies the task's security requirements, including providing a valid ZKP where required. Thus, lower-risk tasks may be assigned to agents with lower trust scores, provided those scores meet the relevant threshold, while higher-risk or critical tasks require agents with correspondingly higher scores. If an agent's score falls below a defined global minimum, a smart contract automatically revokes its VCs. Because access to mTLS connections depends on valid credentials, revocation prevents the agent from establishing authorized connections and effectively removes it from the network.

\section{Influence on MCP}
\label{sec:mcp}

Agent-to-agent communication protocols are not the only protocol in MAS. Agents also rely on the model context protocol (MCP) to gather context from private sources, invoke tools, operate on files, and interact with external software. MCP lets an agent be more autonomous and capable of completing a wider range of tasks. However, with these capabilities come certain  risks. If we want MAS to operate with little human oversight, we have to create a secure agent-to-agent communication protocol to stop potential threats originating from other agents before they even reach the MCP layer. It will not address all of the challenges faced by MCP as some of them originate from the MCP servers themselves.

MCP architecture is composed of three components: MCP server, MCP host and MCP client. The MCP server dictates what an AI model can use, e.g., tools, resources, prompt templates. The MCP host is the AI application in which the model operates and interacts with its user, in our case mostly with upstream agents. It enforces the system boundaries and manages permissions. Often implementing human-in-the-loop checkpoints before any critical execution is done. However, this is not an efficient security solution even in single agent systems, as humans are prone to a phenomenon called alert fatigue, accepting all agentic requests without proper review, which degrades security~\cite{mcpfatigue}. In MAS this mechanism should source information from our dynamic permissions and proofs of intention to create a secure task execution. The last component is the MCP client, which initiates requests to the MCP server on behalf of the MCP host, it also performs sampling, gathers data on used tools and task progress, while also keeping the user updated via notifications~\cite{hou2025modelcontextprotocolmcp}.

MCP servers themselves are at risk of being compromised and used as attack vectors~\cite{hou2025modelcontextprotocolmcp}. This further emphasizes the fact that secure agent-to-agent communication protocol is crucial to be present in such systems. What is also important is that the MCP security solutions work only in properly configured environment, so it should be implemented professionally. Just like in the case of software and docker image repositories, publicly available MCP servers from MCP marketplaces also need cryptographic solutions that validate and authenticate them. On one hand, it helps to identify dangerous MCP servers which are using typo-squatting and impersonation attacks, but still on the other hand if we assume malicious developer types of attack, this solution is not enough. That is where our trust threshold model could help.
Trust thresholding, secure agent-to-agent communication and verified MCP marketplaces help to sift potentially dangerous prompts and actions before they are executed and cause damage. This could together prevent from malicious usage of MCPas .

\section{Semantic Communication}
\label{sec:semmcom}

Sixth-generation networks (6G) are envisioned not merely as faster networks, but as systems tailored to multi-agent systems (MAS) and the internet of everything (IoE), with support for augmented reality (AR), virtual reality (VR), and mixed reality (MR)~\cite{pennanen20256g}. The volume of traffic carried by telecommunications infrastructure is expected to grow exponentially. Moreover, network infrastructure will not simply act as a conduit for data, but as an intelligent, cognitive entity capable of understanding, reasoning, and interacting with users and devices. This vision calls for highly efficient communication to support AI-enabled devices and applications.

Accordingly, the classical information-theoretic paradigm is evolving from Shannon’s non-semantic view of communication toward semantic communication. This shift can be understood through Weaver’s three-level communication framework~\cite{shannon_weaver_1949}: the technical level addresses the effective transmission of messages; the semantic level concerns the meaning of the transmitted data; and the effectiveness level examines how that meaning influences the receiver or affects task execution. Initially developed for human-to-human (H2H) communication, this framework has since been extended to human-to-machine (H2M) and machine-to-machine (M2M) communication. Four use cases for semantic communication in 6G are frequently highlighted: satellite communications, immersive communication, intelligent Internet of Vehicles, and multi-agent AI collaboration~\cite{zhang2026semantic}.

MAS are goal-oriented, they are built to effectively solve different problems. That is why M2M communication focuses more on how we can output the most information while generating little traffic. Semantic communication can significantly help us achieve this. Transmitted information between agents will be specific, crucial for a task. The key will be to effectively process data on edge devices to create messages as meaningful for a specific task as possible, which then will be transferred to some other devices. Their task will be to interpret the messages correctly, and use their meaning effectively. Instead of transferring a lot of data and focusing on retrieving it without errors, we focus on effective communication. Communicating systems are intelligent, understand what they handle, work together to optimize the communication. That creates less traffic in the network, which in turn could decrease latency and energy usage. In addition, semantic communication brings prioritization mechanisms for better management. More resources are given to the more important or time critical data. It also improves the coordination in MAS, letting agents exchange their local states, goals, and summaries.

This helps improve synchronization, workflows, federated learning, and many other processes key for 6G and agentic standards.
Semantic communication can enhance communication security. Rather than exchanging raw data, agents share properties, summaries, and semantic representations. This reduces redundant network traffic, makes sensitive information easier to control, and makes it more difficult to reconstruct the original input. Because communication can be tailored to the context, task, priorities, and data involved, rules can be established to determine the specificity or level of detail in each message. Heavy traffic is undesirable in MAS. Semantic communication can help filter unwanted messages, such as spam and decision noise, thereby improving system reliability. It also enables participants to account for the conversation’s context and goals, making it easier to detect anomalies and unusual requests. This, in turn, makes the system more robust and less prone to errors.

\subsection{Implementation}

Semantic communication enables intelligent processing of text, images, video, and audio, although each modality requires different processing methods and presents distinct challenges~\cite{zhang2026semantic}. At its core, semantic communication uses semantic encoders and decoders that work together to improve interpretation by sharing and building knowledge bases. Semantic encoders extract the underlying meaning from images and other media, compress it into a semantic representation, and transmit the information relevant to the task. This representation is sent over a physical medium through a semantic channel. At the receiver, a semantic decoder reconstructs and interprets the message using the shared context, even if transmission errors have occurred. This approach is effective only when both endpoints share a synchronized knowledge base containing the context, vocabulary, and algorithms needed to interpret the semantic symbols.

Semantic communication is a promising but still developing field, and its practical implementations continue to face structural challenges. One key challenge is maintaining up-to-date, synchronized knowledge bases across heterogeneous, distributed agents. When these knowledge bases are outdated or inconsistent, they can lead to misinterpretations and cascading failures. Another open problem is developing metrics to evaluate transmission quality. In addition, semantic communication relies on deep learning models that can introduce latency and consume substantial energy when deployed on resource-constrained devices. Future research should focus on lightweight, efficient neural architectures for edge computing and on methods for measuring system efficiency. It should also investigate dynamic knowledge-base synchronization protocols and autonomously adaptive systems that can seamlessly translate between different data formats.

\subsection{Standardization}

Semantic communication is considered a crucial part of 6G networks~\cite{zhang2026semantic}. The ITU-T identified semantic communication as an enabler for advanced networks beyond IMT-2020. They initiated technical reports and recommendations, like Y.STR.SemComm to guide global system design. ITU-T is also developing specifications such as F.JSQSUDA for joint semantic query systems. 3GPP actively discusses integration of semantic communication in their future releases. They consider embedding this feature within control channels for quality of service identifiers and semantic-aware channel coding and modulation. Within the realm of MAS, semantic standardization possesses an established history that is currently rapidly evolving to incorporate modern web and artificial intelligence technologies.

The foundational standard in this domain remains the Agent Communication Language developed (ACL) by the FIPA, called FIPA-ACL. It is now formally maintained by an IEEE standards committee. FIPA-ACL provides a rigorous formal semantic framework that explicitly defines both the syntactic structure of messages and their pragmatic effects on the mental states, beliefs, and intentions of agents, thereby enabling highly complex coordination through standardized interaction protocols~\cite{semcomm}.

\section{Discussion}
\label{sec:discussion}

The architectural analysis and valid tests could confirm that the transition from synchronous, point-to-point, deterministic architectures to asynchronous, recursive, zero-trust agent networks renders legacy cryptographic paradigms insufficient. Standard text-based serialization formats like JSON-RPC may not deliver the byte-level determinism mandatory for ZKPs. Furthermore, while traditional transport layer security protects data in transit, it can fail to verify whether an agent hallucinated or maliciously altered the primary user's intent. These vulnerabilities are further compounded by the necessity of migrating to PQC, where the sheer size of standard algorithms like ML-DSA introduces severe handshake bloat. This analysis establishes that cryptographic challenges in MAS communication are multifaceted, spanning identity management, transport security, payload integrity, and intent verification.

To resolve those issues, we hypothesize that multi-agent systems could be based on a framework that actively decouples transport security from the payload integrity. By wrapping CBOR within COSE envelope, the system would achieve the strict byte-level stability. Within the immutable envelope, zk-SNARKs could provide the solution that anchor and continuously verify the semantic intent across the execution chain of agents, without exposing any private witness. Finally, upgrading the transport layer to a crypto-agile, PQC-mTLS optimized with KEMTLS-PDK ensures (for now) quantum resistance, while bypassing the latency overhead from large post-quantum signatures. The resulting integrated architecture provides a practical framework for applying these cryptographic mechanisms to address the identified challenges in MAS communication.

Despite highly theoretical robustness, transitioning from architectural specification to an active implementation may reveal some practical constraints that must be addressed. The primary bottleneck lies in the computational intensity of generating ZKPs. While verification by the receiving agent is highly efficient, the time required to compile an agent's execution trace into an arithmetic circuit and generate the corresponding zk-SNARK could induce some unacceptable delays when coupled with the high latency of LLM inference. Additionally, although KEMTLS mitigates transport layer bloat, the COSE payload faces a scaling challenge known as envelope accumulation. As each agent appends its own proof leading to a linear growth in payload size it can eventually strain the memory limits of edge nodes. Furthermore, the reliance on a blockchain registry to enforce dynamic least privilege introduces a brief vulnerability window tied to block-time finality, during which a revoked but previously authenticated agent might attempt to establish a new connection before the network consensus fully registers its revocation.

To bridge the mentioned gap between the architecture we synthesized and the real-world systems, future research must prioritize the integration of recursive ZKPs. By allowing an agent to generate a single proof that verifies both its own execution and the validity of the preceding agent's proof, the system could maintain a constant envelope size regardless of the execution chain's depth. Furthermore, engineering efforts could focus on developing compilers capable of deterministically translating natural language system prompts into arithmetic circuits without human intervention. It should be emphasized that current zk-SNARKs approaches are based on elliptic curves, so it is important to look for other quantum-proof solutions that we can implement in MAS (like z-STARKs), and then benchmark them. Empirical network simulations are required to check the precise latency reductions achieved by KEMTLS-PDK over standard post-quantum handshakes, ensuring that the theoretical resilience of the architecture scales effectively under heavy concurrent load.

\section{Conclusion}
\label{sec:conclusion}

In this work, we collected and explained novel concepts developing in MAS security and agent-to-agent communication protocols, combining relevant solutions into a unified logical architecture. We discussed what may change in the future, how to sustain protocol robustness, and handle the dynamicity of such systems. Apart from cryptographic challenges, we took into consideration trust thresholds and AI monitoring efforts. However, it is crucial to note, many emerging solutions and architectures may be hard to implement in their entirety or deemed impractical for some use cases. The presented concepts are useful in the research process to inspire, start discussions, and show the general state of multi-agent technologies. Because this study is primarily theoretical, these topics require additional empirical tests and benchmarks, e.g., measuring the overhead of message-level protection with CBOR and COSE. This is especially relevant in order to answer if message-level protection with COSE introduces an acceptable overhead in agent-to-agent communication, particularly with post-quantum signatures. Another crucial part is to measure how does hybrid mTLS performs in MAS? Does it introduce noticeable latencies? Does KEMTLS-PDK resolve this? Is integrating ZKPs a valid approach? Is the proposed architecture scalable? How well does it handle dynamic environments?

\newpage
\printbibliography

@book{shannon_weaver_1949,
  author    = {Shannon, Claude Elwood and Weaver, Warren},
  title     = {The Mathematical Theory of Communication},
  publisher = {University of Illinois Press},
  address   = {Urbana, IL},
  year      = {1949}
}

@online{maswebsite,
  title     = {Multi Agent System in AI},
  author    = {Tanmoy Mishra},
  year      = 2026,
  url       = {https://www.geeksforgeeks.org/artificial-intelligence/multi-agent-system-in-ai/},
  urldate   = {2026-10-12}
}

@online{mcpfatigue,
  title     = {Confirmation Fatigue and the Protocol Gap in Agentic AI Oversight},
  author    = {Changkun Ou},
  year      = 2026,
  url       = {https://changkun.de/blog/ideas/human-in-the-loop-agents/},
  urldate   = {2026-10-12}
}

@online{semcomm,
  title     = {Types of Agent Communication Languages},
  author    = {Alexander de Ridder},
  year      = 2024,
  url       = {https://smythos.com/developers/agent-development/types-of-agent-communication-languages/},
  urldate   = {2026-10-12}
}

@online{google2025a2a,
  author       = {{Google Developers}},
  title        = {A2A: A New Era of Agent Interoperability},
  url          = {https://developers.googleblog.com/en/a2a-a-new-era-of-agent-interoperability/},
  date         = {2025-04-09},
  organization = {Google Developers Blog},
  urldate      = {2026-10-12}
}

@online{mitreAtlas,
    title = {MITRE ATLAS: Adversarial Threat Landscape for Artificial-Intelligence Systems},
    author = {{The MITRE Corporation}},
    year = {2021},
    url = {https://atlas.mitre.org/matrices/ATLAS-matrix},
    urldate = {2026-10-12}
}

@online{owaspAgentic,
    title={OWASP Top 10 for Agentic Applications for 2026},
    author = {{OWASP GenAI Security Project}},
    month = {12},
    year = {2025},
    url = {https://genai.owasp.org/resource/owasp-top-10-for-agentic-applications-for-2026/},
    urldate = {2026-10-12}
}

@book{weiss1999multiagent,
  title={Multiagent Systems: A Modern Approach to Distributed Artificial Intelligence},
  author={Weiss, Gerhard},
  year={1999},
  publisher={MIT Press},
  isbn={978-0-262-23203-6}
}

@misc{hou2025modelcontextprotocolmcp,
    title={Model Context Protocol (MCP): Landscape, Security Threats, and Future Research Directions}, 
    author={Xinyi Hou and Yanjie Zhao and Shenao Wang and Haoyu Wang},
    year={2025},
    eprint={2503.23278},
    archivePrefix={arXiv},
    primaryClass={cs.CR}
}

@misc{touheed2026sszkr,
    title={SS-ZKR: Spatial-Semantic Zero-Knowledge Routing for Privacy-Preserving Multi-Agent Collaboration}, 
    author={Hassan Touheed},
    year={2026},
    eprint={2606.00962},
    archivePrefix={arXiv},
    primaryClass={cs.CR}
}

@inproceedings{zheng2025agentsight,
    author = {Zheng, Yusheng and Hu, Yanpeng and Yu, Tong and Quinn, Andi},
    title = {AgentSight: System-Level Observability for AI Agents Using eBPF},
    year = {2025},
    isbn = {9798400722059},
    publisher = {Association for Computing Machinery},
    address = {New York, NY, USA},
    doi = {10.1145/3766882.3767169},
    booktitle = {Proceedings of the 4th Workshop on Practical Adoption Challenges of ML for Systems},
    pages = {110–115},
    numpages = {6},
    location = {Seoul, Republic of Korea},
    series = {PACMI '25}
}

@inproceedings{schwabe2020pqtls,
    author = {Schwabe, Peter and Stebila, Douglas and Wiggers, Thom},
    title = {Post-Quantum TLS Without Handshake Signatures},
    year = {2020},
    isbn = {9781450370899},
    publisher = {Association for Computing Machinery},
    address = {New York, NY, USA},
    doi = {10.1145/3372297.3423350},
    booktitle = {Proceedings of the 2020 ACM SIGSAC Conference on Computer and Communications Security},
    pages = {1461–1480},
    numpages = {20},
    location = {Virtual Event, USA},
    series = {CCS '20}
}

@inproceedings{nirban2026zerotrust,
    author="Nirban, Harshvardhan Singh
    and Hinduja, Bharat
    and Mangal, Bhaskar
    and Chakka, Venugopal
    and Sharma, Yashvardhan
    and Srivastava, Sharanya Rajan
    and Srivastava, Shaurya Rajan
    and Vaidya, Rohan
    and Srivastava, Sumit Rajan
    and Bhatia, Ashutosh",
    editor="Barolli, Leonard
    and K. G.  Seah, Winston
    and Woungang, Isaac",
    title="A Zero-Trust Identity Framework for Autonomous AI Agents: Cryptographic Attestation and Policy-Driven Authorization",
    booktitle="Advanced Information Networking and Applications",
    year="2026",
    publisher="Springer Nature Switzerland",
    address="Cham",
    pages="258--269",
    isbn="978-3-032-23311-0",
    doi="10.1007/978-3-032-23311-0_22"
}

@phdthesis{wiggers2024pqtls,
    author = {Thom Wiggers},
    title = {Post-Quantum TLS},
    school = {Radboud University},
    address = {Nijmegen, The Netherlands},
    date = {2024-01-09},
    url = {https://thomwiggers.nl/publications/thesis/thesis.pdf},
}

@inproceedings{huang2026zerotrust,
    author={Huang, Ken and Narajala, Vineeth Sai and Yeoh, John and Ross, Jason and Lambe, Mahesh and Raskar, Ramesh and Harkati, Youssef and Huang, Jerry and Habler, Idan and Hughes, Chris},
    booktitle={2026 International Conference on AI x Data and Knowledge Engineering (AIxDKE)}, 
    title={A Novel Zero-Trust Identity Framework for Agentic AI: Decentralized Authentication and Fine-Grained Access Control}, 
    year={2026},
    volume={},
    number={},
    pages={98-101},
    doi={10.1109/AIxDKE67294.2026.00024}
}

@inproceedings{schemitt2025pqc,
    author={Schemitt, Alison Gonçalves and da Silva, Henrique Fan and Lunardi, Roben Castagna and Kreutz, Diego and Mansilha, Rodrigo Brandão and Zorzo, Avelino Francisco},
    booktitle={2025 IEEE 24th International Conference on Trust, Security and Privacy in Computing and Communications (TrustCom)}, 
    title={Assessing the Impact of Post-Quantum Digital Signature Algorithms on Blockchains}, 
    year={2025},
    volume={},
    number={},
    pages={2373-2380},
    doi={10.1109/Trustcom66490.2025.00276}
}

@misc{zou2025blocka2a,
    title={BlockA2A: Towards Secure and Verifiable Agent-to-Agent Interoperability}, 
    author={Zhenhua Zou and Zhuotao Liu and Lepeng Zhao and Qiuyang Zhan},
    year={2025},
    eprint={2508.01332},
    archivePrefix={arXiv},
    primaryClass={cs.CR}
}

@misc{rfc8785,
    series =    {Request for Comments},
    number =    8785,
    howpublished =  {RFC 8785},
    publisher = {RFC Editor},
    doi =       {10.17487/RFC8785},
    author =    {A. Rundgren and B. Jordan and S. Erdtman},
    title =     {{JSON Canonicalization Scheme (JCS)}},
    pagetotal = 20,
    year =      2020,
    month =     jun,
}

@misc{rfc9954,
    series =    {Request for Comments},
    number =    9954,
    howpublished =  {RFC 9954},
    publisher = {RFC Editor},
    doi =       {10.17487/RFC9954},
    author =    {D. Stebila and S. Fluhrer and S. Gueron},
    title =     {{Hybrid Key Exchange in TLS 1.3}},
    pagetotal = 18,
    year =      2026,
    month =     jul,
}

@misc{rfc9052,
    series =    {Request for Comments},
    number =    9052,
    howpublished =  {RFC 9052},
    publisher = {RFC Editor},
    doi =       {10.17487/RFC9052},
    author =    {J. Schaad},
    title =     {{CBOR Object Signing and Encryption (COSE): Structures and Process}},
    pagetotal = 66,
    year =      2022,
    month =     aug,
}

@misc{dewitt2026open,
    title={Open Challenges in Multi-Agent Security: Towards Secure Systems of Interacting AI Agents}, 
    author={Christian Schroeder de Witt and Klaudia Krawiecka and Igor Krawczuk and Ben Hagag and William L. Anderson and Peter Belcak and Ben Bucknall and Xiaohong Cai and Ayush Chopra and Doron Cohen and Ron F. Del Rosario and Andis Draguns and Annie Gray and Keren Katz and Vasilios Mavroudis and Jaron Mink and Sumeet Ramesh Motwani and Jonathan Petit and Leif-Sebastian Rembeck and Chandler Smith and John Sotiropoulos and Steven Young and Sarah Scheffler and Mary Llewellyn},
    year={2026},
    eprint={2505.02077},
    archivePrefix={arXiv},
    primaryClass={cs.CR}
}

@article{karim2025aiagents,
    AUTHOR = {Karim, Md Monjurul and Van, Dong Hoang and Khan, Sangeen and Qu, Qiang and Kholodov, Yaroslav},
    TITLE = {AI Agents Meet Blockchain: A Survey on Secure and Scalable Collaboration for Multi-Agents},
    JOURNAL = {Future Internet},
    VOLUME = {17},
    YEAR = {2025},
    NUMBER = {2},
    ARTICLE-NUMBER = {57},
    ISSN = {1999-5903},
    DOI = {10.3390/fi17020057}
}

@article{saltzer1984endtoend,
    author = {Saltzer, J. H. and Reed, D. P. and Clark, D. D.},
    title = {End-to-end arguments in system design},
    year = {1984},
    issue_date = {Nov. 1984},
    publisher = {Association for Computing Machinery},
    address = {New York, NY, USA},
    volume = {2},
    number = {4},
    issn = {0734-2071},
    doi = {10.1145/357401.357402},
    journal = {ACM Trans. Comput. Syst.},
    month = {11},
    pages = {277–288},
    numpages = {12},
}

@article{neulinger2025fostering,
    author    = {Neulinger, Alexander and Sparer, Lukas},
    title     = {Fostering AI alignment through blockchain, proof of personhood and zero knowledge proofs},
    journal   = {Cluster Computing},
    year      = {2025},
    volume    = {28},
    number    = {15},
    pages     = {983},
    month     = {10},
    issn      = {1573-7543},
    doi = {10.1007/s10586-025-05729-8}
}

@inbook{huang2025agentic,
    author="Huang, Ken
    and Hughes, Chris",
    title="Agentic AI Communication Protocols and Security",
    bookTitle="Securing AI Agents: Foundations, Frameworks, and Real-World Deployment",
    year="2025",
    publisher="Springer Nature Switzerland",
    address="Cham",
    pages="81--110",
    isbn="978-3-032-02130-4",
    doi="10.1007/978-3-032-02130-4_4"
}

@inproceedings{li2026a2asecbench,
    author = {Li, Tianhao and Chu, Chuangxin and Zheng, Yujia and Zhang, Bohan and Gong, Neil and Xiao, Chaowei},
    booktitle = {International Conference on Learning Representations},
    editor = {C. Vondrick and B. Hariharan and C. Raffel and L. Pinto and D. Yang and A. Faust},
    pages = {122168--122191},
    title = {A2ASecBench: A Protocol-Aware Security Benchmark for Agent-to-Agent Multi-Agent Systems},
    volume = {2026},
    year = {2026}
}

@article{louck2026security,
    author = {Louck, Yedidel and Dvir, Amit and Stulman, Ariel},
    title = {Security Analysis of Agentic AI Communication Protocols: A Comparative Evaluation},
    year = {2026},
    publisher = {Association for Computing Machinery},
    address = {New York, NY, USA},
    doi = {10.1145/3803431},
    journal = {ACM Trans. AI Secur. Priv.},
    month = {3}
}

@article{zhang2026semantic,
    author={Zhang, Ping and Xu, Xiaodong and Sun, Mengying and Gao, Haixiao and Ma, Nan and Wang, Xiaoyun and Zhang, Ruichen and Wang, Jiacheng and Niyato, Dusit},
    journal={IEEE Communications Standards Magazine}, 
    title={Toward Native AI in 6G Standardization: The Roadmap of Semantic Communication}, 
    year={2026},
    volume={},
    number={},
    pages={1-11},
    doi={10.1109/MCOMSTD.2026.3669900}
}

@article{pennanen20256g,
    author={Pennanen, Harri and Hänninen, Tuomo and Tervo, Oskari and Tölli, Antti and Latva-Aho, Matti},
    journal={IEEE Access}, 
    title={6G: The Intelligent Network of Everything}, 
    year={2025},
    volume={13},
    number={},
    pages={1319-1421},
    doi={10.1109/ACCESS.2024.3521579}
}

@article{ferrag2026agents,
    author={Ferrag, Mohamed Amine and Lakas, Abderrahmane and Debbah, Mérouane},
    journal={IEEE Open Journal of the Communications Society}, 
    title={6G Needs Agents: Toward Agentic AI-Native Networks for Autonomous Intelligence}, 
    year={2026},
    volume={7},
    number={},
    pages={7254-7282},
    doi={10.1109/OJCOMS.2026.3707904}
}

\end{document}